\PassOptionsToPackage{sort&compress}{natbib}
\documentclass[final,5p,times,twocolumn,authoryear]{elsarticle}

\usepackage{amsmath,amssymb}
\usepackage{lipsum}
\usepackage{graphicx}
\usepackage{bm}
\usepackage{xcolor}
\usepackage[colorlinks=true,linkcolor=blue,citecolor=blue,urlcolor=blue]{hyperref}
\setcitestyle{numbers,square,comma}

\journal{Physics Letters B}

\begin{document}

\begin{frontmatter}

\title{Polarization of strongly lensed gravitational waves as a probe of the lens-induced orientations of electromagnetic counterparts}

\author[first]{Kyungmin Kim}
\ead{kkim@kasi.re.kr}
\affiliation[first]{organization={Korea Astronomy and Space Science Institute},
            addressline={776 Daedeok-daero, Yuseong-gu},
            city={Daejeon},
            postcode={34055},
            country={Republic of Korea}}

\begin{abstract}
Strong gravitational lensing is a well-established phenomenon in electromagnetic (EM) observations, where the lensed images and arcs are resolved on the two-dimensional sky and constrain the mass distribution of the lens. Strongly lensed gravitational waves (GWs) are also expected to be detected routinely by next-generation detectors, but their multiple images are recorded only as one-dimensional time series without positions. Among them, lensed binary neutron star mergers can be accompanied by EM counterparts with resolvable structures, such as relativistic jets. In EM lensing, the lens-induced change in the apparent orientation of such a structure can be measured only relative to other images or statistically, because its intrinsic orientation is not directly observable. In this work, we show that the GW polarization, which the lens leaves unchanged, provides this missing reference, i.e., a compass, so that the orientation offset of each image can be measured in absolute terms. We place this orientation layer on the EM image plane together with image positions from the lensed host arcs and Fermat-potential heights from GW time delays and magnifications. Using a mock population of such mergers observable by the Einstein Telescope and Cosmic Explorer, we find that the predicted image positions and arrival times are calibrated once the detection selection is included. The lens-induced orientation offsets reach tens of degrees and are detectable in up to a third of the events, at viewing angles above about $40^\circ$ for quadrupole-dominated signals.
\end{abstract}


\end{frontmatter}

%% =====================================================================
\section{Introduction}
\label{sec:intro}

Strong gravitational lensing by galaxies, as seen in electromagnetic (EM) observations, constrains the lens mass distribution through resolved images and arcs, and it measures cosmological distances through time delays~\cite{Schneider1992,Treu2010}; the image positions and time delays are set by the stationary points of the Fermat potential and by its values there, respectively~\cite{Blandford1986}. Similar to light, gravitational waves (GWs) are lensed by the same potentials~\cite{Ohanian1974,Deguchi1986,Wang1996,Nakamura1998,Takahashi2003}, and in geometric optics a strongly lensed GW arrives as multiple images, i.e., copies of the signal that differ in arrival time, amplitude, and Morse phase~\cite{Dai2017,Ezquiaga2021}, but without positions.

Since galaxy-scale lensing affects only a small fraction of detectable mergers~\cite{Ng2018,Li2018,Oguri2018}, no lensed event has been identified with high confidence in the LIGO--Virgo--KAGRA searches~\cite{Hannuksela2019,LVK2021lensing,LVK2023lensing,LVK2025lensing}; GW231123 showed the strongest, but inconclusive, support~\cite{LVK2025lensing}. For the next-generation detectors Einstein Telescope (ET)~\cite{Punturo2010,Hild2011,Maggiore2020,Branchesi2023} and Cosmic Explorer (CE)~\cite{Reitze2019,Abbott2017CE,Evans2021}, tens to hundreds of strongly lensed events per year are anticipated~\cite{Piorkowska2013,Biesiada2014,Ding2015,Li2018,Wierda2021,Xu2022,ETScience2026}, including tens of lensed binary neutron star (BNS) mergers among the $\sim10^5$ BNS detections per year~\cite{Ronchini2022,Branchesi2023,ETScience2026}.

However, GW detectors cannot resolve arcsecond image separations, with a sky localization of at best $\sim10^{-3}\,{\rm deg}^2$~\cite{Chen2025}; instead, the relative arrival times and magnification ratios, combined with a lens model, constrain where the images lie~\cite{Pagano2020}. Additional images improve the localization of LISA binaries~\cite{Seto2004}, and \cite{Hannuksela2020,Yu2020} showed that time delays and magnification ratios can identify the lensed host and localize the merger to sub-arcsecond precision, reaching $\sim10$\,mas for bright systems~\cite{Wempe2024}.

Lensed GWs with EM counterparts can also probe cosmology and GW propagation~\cite{Liao2017,Collett2017,Fan2017,Yang2019,Smith2025}. Unlike binary black holes, lensed BNS mergers can be accompanied by kilonova, gamma-ray burst, and afterglow counterparts, whose lensed rates and follow-up strategies have been studied~\cite{ChenLuZeng2026,Ryczanowski2025,Magare2023}. Such counterparts can be spatially resolved, e.g., the jet motion of GW170817 with very long baseline interferometry (VLBI)~\cite{Mooley2018,Ghirlanda2019,Hotokezaka2019}, so that a lensed BNS counterpart can provide both the position and the apparent orientation of a structure in each image.

Despite these efforts, the GW polarization has not been considered in this joint picture, although its polarization angle $\psi$ is the position angle of the projected orbital angular momentum on the image plane and is parallel-transported by static lenses with negligible rotation~\cite{Hou2019}. Lensed GWs have been proposed to test the polarization content of gravity~\cite{Goyal2021,MaganaHernandez2022} and helicity-dependent propagation~\cite{Andersson2023}, and the afterglow polarization of GW170817~\cite{GW170817,GW170817MMA} has been used as a prior in such tests~\cite{Imam2026}, but none relates the GW polarization to the image-plane structure of a lensed counterpart.

In lensing, the intrinsic orientation of a source, which is not directly observable, is a fundamental limitation, i.e., the ``shape noise'' of weak lensing~\cite{Bartelmann2001}; in strong lensing, it allows only the relative orientations between images to be measured. In radio astronomy, the unlensed polarization of light has been used as a proxy for the intrinsic orientation~\cite{Surpi1999,Brown2011,Francfort2021}, but only statistically because of its intrinsic misalignment with the morphology. In a BNS merger, by contrast, both $\psi$ and the jet axis are tied to the orbital angular momentum. Therefore, the GW polarization can serve as a physical reference for the intrinsic orientation of the jet, and its offset from the apparent jet axis measures the lens-induced orientation offset of each image in absolute terms. Motivated by this, in this work, we propose the GW polarization as an orientation reference for strong lensing, place it on the EM image plane together with the image positions and time delays (Sec.~\ref{sec:framework}), and validate it with mock lensed BNS mergers (Secs.~\ref{sec:validation} and~\ref{sec:results}). To our knowledge, this is the first work to represent the full set of GW lensing observables, including the polarization, on the EM image plane, and to use the GW polarization as a reference for the orientations of lensed EM counterparts.

%% =====================================================================
\section{Image-plane framework}
\label{sec:framework}

In the thin-lens approximation, the image-plane position $\bm\theta=(\theta_1,\theta_2)$ and the source-plane position $\bm\beta$ obey $\bm\beta=\bm\theta-\nabla\Psi(\bm\theta)$~\cite{Schneider1992}, where $\Psi$ is the lensing potential in angular units. Image $j$ forms at $\bm\theta_j$ and arrives at
\begin{equation}
t_j=\frac{D_{\Delta t}}{c}\,\tau(\bm\theta_j;\bm\beta),\qquad
\tau(\bm\theta;\bm\beta)=\tfrac12|\bm\theta-\bm\beta|^2-\Psi(\bm\theta),
\label{eq:fermat}
\end{equation}
where $\tau$ is the Fermat potential, $c$ is the speed of light, and $D_{\Delta t}=(1+z_l)D_lD_s/D_{ls}$ is the time-delay distance, with lens redshift $z_l$ and angular-diameter distances $D_l$, $D_s$, and $D_{ls}$ to the lens, to the source, and between them. Images form at the stationary points of $\tau$~\cite{Blandford1986}, and the local mapping is described by the Jacobian $A_j=\partial\bm\beta/\partial\bm\theta$ at $\bm\theta_j$, whose isotropic and traceless parts define the convergence $\kappa=\tfrac12\nabla^2\Psi$ and the shear of amplitude $\gamma$~\cite{Schneider1992}. The magnification is $\mu_j=1/\det A_j$, whose sign is the image parity, and minima, saddles, and maxima of $\tau$ are type I, II, and III images with Morse indices $n_j=0,1/2,1$. Each image carries a copy of the unlensed signal $\tilde h(f)$~\cite{Dai2017,Ezquiaga2021},
\begin{equation}
\tilde h_j(f)=|\mu_j|^{1/2}\exp\!\left[2\pi i f t_j-i\pi n_j\,\mathrm{sgn}(f)\right]\tilde h(f),
\label{eq:lensedwave}
\end{equation}
where $f$ is the GW frequency. Throughout this work, image $j$ is the lensed image at $\bm\theta_j$, recorded by GW detectors as a separate signal (``GW image'' $j$) without position; its position is thus inferred only through the lens model.

\subsection{Three layers on the image plane}
\label{sec:layers}

We organize the observables of a lensed GW with an EM counterpart into three layers. First, the \emph{position} layer comes from EM imaging: the lensed host arcs constrain the source-plane surface brightness $S(\bm\beta)$ of the host and the lens model, i.e., a set of parameters $M$ (e.g., the Einstein radius $\theta_E$, density slope, and external shear) that specifies $\Psi$; for brevity, we express this lens-model dependence through $\Psi$ hereafter. The merger position $\bm\beta_{\rm GW}$ then has the prior
\begin{equation}
p(\bm\beta_{\rm GW}\mid S)\propto S(\bm\beta_{\rm GW})\,p_{\rm off}(\bm\beta_{\rm GW}),
\label{eq:posprior}
\end{equation}
where $p_{\rm off}$ accounts for displacements of mergers from the stellar light of the host, e.g., due to natal kicks and long delay times, and a constant $p_{\rm off}$ corresponds to mergers that trace the host light~\cite{Hannuksela2020,Yu2020,Chen2025}. Second, the \emph{height} layer comes from GW data: time delays constrain relative values of $\tau$ for a given $D_{\Delta t}$, and magnification ratios constrain its local curvature. Given an independent source redshift $z_s$ and a cosmology, the observed distance of each image, $D_L^{{\rm obs},j}=D_L(z_s)|\mu_j|^{-1/2}$, also gives its absolute magnification; for BNS mergers, $z_s$ follows from the tidal signature for a given equation of state~\cite{Messenger2012,DelPozzo2017}. Finally, the \emph{orientation} layer combines $\psi$ with the axes of resolved EM structures. Since a quadrupole-only signal determines $\psi$ only modulo $90^\circ$ (Sec.~\ref{sec:asymmetry}), we represent it by a cross-shaped ``glyph'' with $90^\circ$ symmetry, common to all images because static lenses parallel-transport the polarization~\cite{Hou2019}.

\begin{figure*}[t]
\centering
\includegraphics[width=1\textwidth]{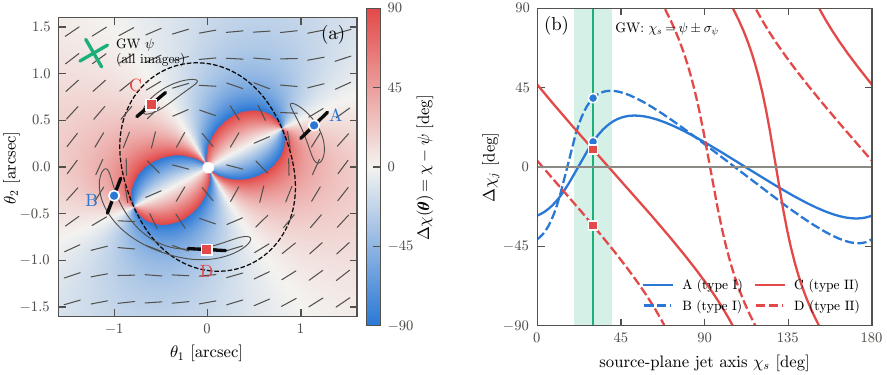}
\caption{The polarization--Jacobian asymmetry for an illustrative quadruple lens, i.e., a singular isothermal sphere with Einstein radius $\theta_E=1\,{\rm arcsec}$ and external shear $\gamma=0.12$ at $z_l=0.5$ and $z_s=2$. (a)~Image-plane offset $\Delta\chi(\bm\theta)=\chi(\bm\theta)-\psi$ for a source-plane jet axis $\chi_s=\psi=30^\circ$, calculated from Eqs.~\eqref{eq:chimap} and~\eqref{eq:dchi} with $\epsilon=0$. Gray bars indicate the apparent jet axis at each position, and black bars indicate the axes at the four image positions. The green cross at the top left is the GW glyph, whose two perpendicular bars lie along $\psi$ and $\psi+90^\circ$; it has the same orientation in every image. The thin contour outlines the lensed host, and the dashed curve marks the critical curve. (b)~Offset for each image as a function of $\chi_s$. Assuming jet--orbit alignment, the GW polarization constrains $\chi_s=\psi$; the vertical line marks the reference value, and the band shows the joint ET+CE $\pm1\sigma$ uncertainty at an inclination of $\iota=50^\circ$. Differences between the image-axis angles can be measured without the GW reference. The apparent axes of saddle images (red) rotate in the opposite direction to the source-plane axis.}
\label{fig:orientation}
\end{figure*}

\subsection{The polarization--Jacobian asymmetry}
\label{sec:asymmetry}

Lensing maps a source-plane direction $\hat{\bm e}(\chi_s)=(\cos\chi_s,\sin\chi_s)$ into $A_j^{-1}\hat{\bm e}(\chi_s)$ in image $j$~\cite{Schneider1992}, so that the axis of a resolved structure such as a jet, with position angle $\chi_s$ on the source plane, appears in image $j$ at
\begin{equation}
\chi_j=\arg\!\left[A_j^{-1}\hat{\bm e}(\chi_s)\right]\pmod{\pi},
\label{eq:chimap}
\end{equation}
where $\arg[\bm v]$ is the position angle of $\bm v$ and all angles are measured from the $\theta_1$ axis. This mapping depends on the lens only through the reduced shear at image $j$, i.e., its direction and amplitude $g_j=\gamma_j/(1-\kappa_j)$~\cite{Schneider1992}, and it underlies the polarization--morphology tests of radio lensing~\cite{Surpi1999,Francfort2021}. Writing $\chi_s=\psi+\epsilon$ with a jet--orbit misalignment $\epsilon$, we define the per-image offset from the GW polarization angle,
\begin{equation}
\Delta\chi_j\equiv\chi_j-\psi\pmod{\pi},
\label{eq:dchi}
\end{equation}
which equals $\epsilon$ in the absence of lensing. Since $|g_j|\sim1$ in strong lensing, we evaluate Eq.~\eqref{eq:chimap} exactly rather than in the weak-shear limit. For an illustrative quadruple lens, the apparent axis is offset by tens of degrees along most of the arcs, and the axes of saddle images counter-rotate as $\chi_s$ varies (Fig.~\ref{fig:orientation}).

The role of the GW is to fix $\chi_s$: without the GW, only the $N-1$ differences between the EM axes of $N$ images are measurable, whereas with the GW, $\chi_s$ is known up to $\epsilon$, so that all $N$ offsets, including the offset common to all images, constrain the Jacobians. Note that a quadrupole-only signal determines $\psi$ only modulo $90^\circ$, because $\psi\to\psi+\pi/2$ is absorbed by the coalescence phase. We assume that the correct mode is selected by mapping the observed axes back to the source plane, $\chi_s=\arg[A_j\hat{\bm e}(\chi_j)]$, which is reliable for $|\epsilon|$ and uncertainties well below $45^\circ$, and quote $\sigma_\psi$ within that mode.

Treating $\psi$, the EM axes, and $\epsilon$, a predictive scatter about alignment with standard deviation $\sigma_\epsilon$, as independent Gaussian quantities, the offset uncertainty of image $j$ is
\begin{equation}
\sigma_{\Delta\chi,j}^2=\sigma_\psi^2+\sigma_{\chi,j}^2+\sigma_\epsilon^2,
\label{eq:sigma}
\end{equation}
where $\sigma_\psi$ is from the joint fit and $\sigma_{\chi,j}$ is the EM-axis uncertainty of image $j$. We define the per-image significance $\mathcal{S}_j=|\Delta\chi_j|/\sigma_{\Delta\chi,j}$, i.e., the offset in units of its standard deviation under the null hypothesis of no lens-induced offset, and regard an event as a detection if $\max_j\mathcal{S}_j\geq3$, i.e., a $3\sigma$ deviation in at least one image. Under the null hypothesis, this corresponds to a false-alarm probability of 0.27\% per image and at most about 1\% per event for up to four images (Bonferroni bound)~\cite{Bayer2020,Caliskan2023luck}; this bound holds regardless of the correlation between images induced by the shared $\psi$ and $\epsilon$, which we do not exploit. For the common offset $\overline{\Delta\chi}=N^{-1}\sum_j\Delta\chi_j$, which EM differences cannot provide, we use
\begin{equation}
\mathcal{S}_{\rm cm}=\frac{|\overline{\Delta\chi}|}{\left(\sigma_\psi^2+\sigma_\epsilon^2+\sigma_\chi^2/N\right)^{1/2}},
\label{eq:scm}
\end{equation}
for a common EM-axis uncertainty $\sigma_\chi$.

Finally, a single lens model links the three layers in the joint likelihood of the EM data $D_{\rm EM}$ (arc imaging and counterpart astrometry) and the GW strain $d_j$ of each image,
\begin{equation}
\begin{split}
\mathcal{L}={}&\mathcal{L}_{\rm EM}(D_{\rm EM}\mid \Psi,S,\bm\beta_{\rm GW})\\
&\times\prod_j\Bigl[\mathcal{L}_{\rm GW}(d_j\mid\bm\vartheta,|\mu_j|,t_j,n_j)\\
&\qquad\quad\times\mathcal{L}_{\rm str}(\chi_j^{\rm obs}\mid A_j,\psi,\epsilon)\Bigr],
\end{split}
\label{eq:likelihood}
\end{equation}
where $\bm\vartheta$ denotes the GW source parameters shared by all images, including $\psi$, $\chi_j^{\rm obs}$ is the measured structure axis, and $|\mu_j|$, $t_j$, $n_j$, and $A_j$ are functions of $(\Psi,\bm\beta_{\rm GW})$ and the cosmology.

%% =====================================================================
\section{Mock lensed BNS mergers}
\label{sec:validation}

We consider lensed BNS mergers whose EM counterparts, such as kilonovae, reveal the true image positions and arrival times. We simulate a population of galaxy-scale lenses, host galaxies, and BNS sources, and select the systems with at least two images detectable by ET and CE (Sec.~\ref{sec:population}). For each selected system, we then estimate how precisely the GW data measure the source redshift, polarization angle, and image amplitudes and arrival times with a Fisher analysis in which all images share the source parameters (Sec.~\ref{sec:gwanalysis}).

\begin{figure*}[t]
\centering
\includegraphics[width=1\textwidth]{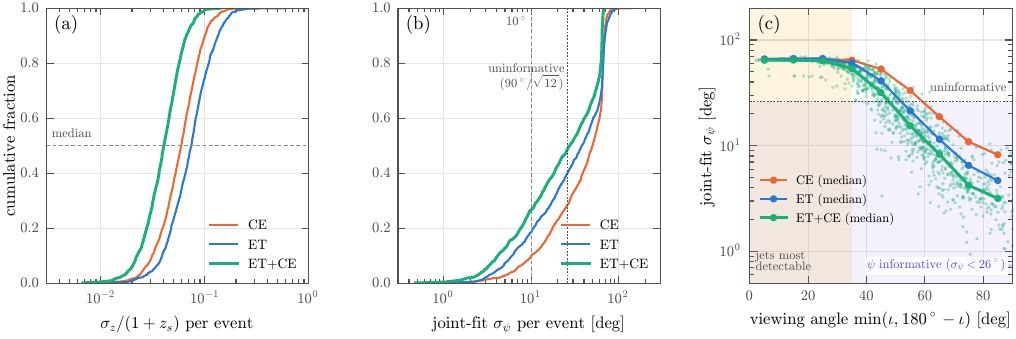}
\caption{GW measurement precision across the mock population for ET, CE, and ET+CE. (a)~Cumulative distribution of the per-event $1\sigma$ uncertainty in $z_s$, divided by $1+z_s$, from the tidal terms; the dashed line marks the median. (b)~Cumulative distribution of the joint-fit $1\sigma$ uncertainty in $\psi$; the vertical lines mark $10^\circ$ and the uninformative value $90^\circ/\sqrt{12}\simeq26^\circ$ for $\psi$ defined modulo $90^\circ$. (c)~Uncertainty in $\psi$ as a function of viewing angle for ET+CE (points), with running medians for ET, CE, and ET+CE. The horizontal dotted line marks the uninformative value, below which $\psi$ is measured (violet region), and the yellow band indicates the viewing angles at which jets are most easily detected. No event lies in their overlap, i.e., $\sigma_\psi<26^\circ$ only at viewing angles above $40^\circ$, which limits test V3 (Sec.~\ref{sec:v3}).}
\label{fig:gw}
\end{figure*}

\subsection{Lens and source population}
\label{sec:population}

We adopt a flat $\Lambda$CDM cosmology with $H_0=67.7\,{\rm km\,s^{-1}\,Mpc^{-1}}$ and $\Omega_m=0.31$, consistent with the Planck 2018 results~\cite{Planck2018}. We draw $4\times10^4$ lensed BNS candidates with $z_s\leq4$ from $p(z_s)\propto\mathcal{R}(z_s)(1+z_s)^{-1}(dV_c/dz_s)D_c^3$, where $\mathcal{R}$ is the star-formation history~\cite{Madau2014} used without merger delays, $(1+z_s)^{-1}$ converts to the observer frame, $V_c$ is the comoving volume, and $D_c^3$ approximates the scaling of the optical depth with the comoving distance $D_c$~\cite{Turner1984}. The lens redshifts follow the isothermal optical-depth kernel $\propto(D_{c,l}D_{c,ls}/D_{c,s})^2dD_{c,l}/dz_l$.

The lenses are spherical power laws, $\rho\propto r^{-\gamma'}$ with $\gamma'\sim\mathcal{N}(2,0.1)$, where $\mathcal{N}(a,b)$ denotes a Gaussian of mean $a$ and standard deviation $b$, plus shear with a Rayleigh scale of 0.08 that also mimics ellipticity. Velocity dispersions follow~\cite{Choi2007} weighted by $\sigma_v^4$ and set the Einstein radius $\theta_E=4\pi(\sigma_v/c)^2D_{ls}/D_s$. Hosts are exponential disks with a median effective radius of $0.2\,{\rm arcsec}$ that mergers trace. Neutron-star masses follow $\mathcal{N}(1.33,0.09)\,M_\odot$~\cite{Ozel2012}, and tidal deformabilities follow $\Lambda(m)=300\,(m/1.4\,M_\odot)^{-6}$, consistent with GW170817~\cite{LVC2018EOS} and an approximately mass-independent radius~\cite{De2018}.

We find the images with a semi-analytic solver and retain systems with at least two GW images (``detected images'' hereafter) above a signal-to-noise ratio of 8 in the joint ET and CE network (ET+CE hereafter), $\theta_E\geq0.2\,{\rm arcsec}$, and a maximum image separation $\geq0.3\,{\rm arcsec}$. The 5908 selected systems have median $z_s=2.3$ and $\theta_E=0.63\,{\rm arcsec}$, and the median of the longest delay between detected images is 48~days. Counting detected images, 93\%, 5\%, and 2\% are doubles, quadruples, and triples, respectively.

\subsection{GW analysis}
\label{sec:gwanalysis}

We use the \textsc{bilby}~\cite{Ashton2019} noise curves of ET-D~\cite{Hild2011} for three $60^\circ$ interferometers in Sardinia and of CE~\cite{Abbott2017CE} for a 40\,km detector at a Hanford-like site, with four times that noise for a 20\,km detector at a Livingston-like site. We use a nonspinning 3.5 post-Newtonian TaylorF2 waveform~\cite{Arun2005} with tidal terms~\cite{Flanagan2008,Wade2014}, which constrain $z_s$ for a fixed equation of state~\cite{Messenger2012,DelPozzo2017}, and antenna patterns that include the Earth's rotation.

For each event, a joint Fisher matrix shares the redshifted chirp mass, symmetric mass ratio, $z_s$, inclination $\iota$, and $\psi$ across images, whereas each image has its own distance $D_L(z_s)|\mu_j|^{-1/2}$, coalescence time, and phase $\phi_{c,j}$. Since $\phi_{c,j}$ absorbs the Morse phase, the image types are taken from the lens model.

ET+CE yields a median $\sigma_z/(1+z_s)=0.04$ and $\sigma_\psi<10^\circ$ for 27\% of the events, compared with 18\% and 10\% for ET and CE alone (Figs.~\ref{fig:gw}(a) and (b)). Hence, GW-only absolute magnifications are limited by the distance--inclination degeneracy rather than by the redshift (Sec.~\ref{sec:v2}), whereas $\psi$, the orientation reference, is measured only at viewing angles above about $40^\circ$, outside the range in which jets are most easily detected (Fig.~\ref{fig:gw}(c)).

\begin{figure*}[!tp]
\centering
\includegraphics[width=\textwidth]{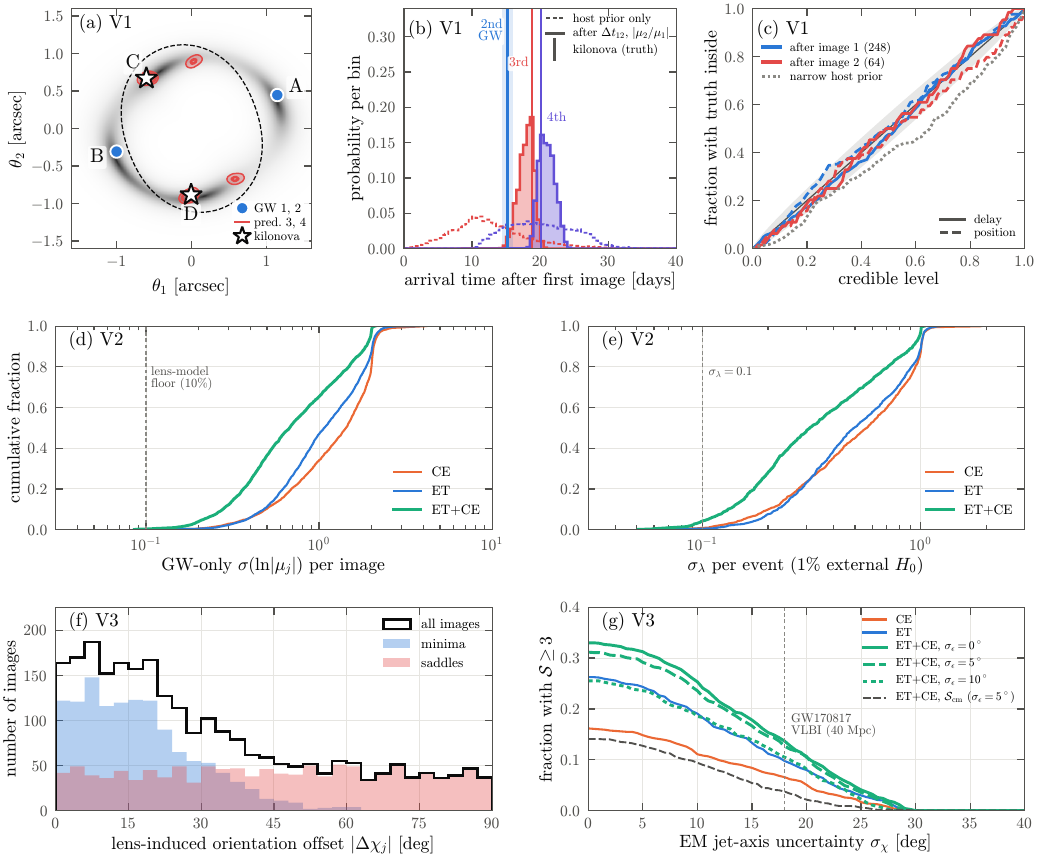}
\caption{Results of tests V1 (top), V2 (middle), and V3 (bottom). Panels (a) and (b) show an illustrative lens with a fixed lens model. (a)~Predicted positions of the third and fourth images after detection of the second GW image ($68\%$ and $95\%$ credible regions, red), overlaid on the lensed host light used to construct the prior. Circles mark the positions of the first two images, and stars mark the true positions of the later images; the off-arc regions correspond to a mirror configuration. (b)~Predicted arrival times relative to the first image, using the host prior alone (dotted) and including the second GW image (filled). Vertical lines mark the true arrival times; the predictions are $17.9\pm1.1$ and $21.0\pm1.1$ days for the true values of 18.8 and 20.1 days, and 77\% of the positional posterior lies around the true configuration. (c)~Coverage across the mock population, including lens-model uncertainty, after the first GW image (blue) and the second (red), for time delays (solid) and positions (dashed). The shaded band shows the 95\% binomial scatter. The gray dotted curve shows the coverage obtained with a host prior that is too narrow. (d)~Cumulative distribution of the GW-only $1\sigma$ uncertainty in $\ln|\mu_j|$ of each detected image for ET, CE, and ET+CE; the dashed line marks the 10\% lens-model uncertainty floor. (e)~Per-event $1\sigma$ uncertainty in the mass-sheet parameter $\lambda$, obtained by combining all images and including the lens-model uncertainty floor and a 1\% external prior on $H_0$. (f)~Distribution of the lens-induced orientation offsets $|\Delta\chi_j|$ of the detected images for $\epsilon=0$, for all images and separately for minima and saddle points. (g)~Fraction of events with at least one image satisfying $\mathcal{S}_j\geq3$ as a function of the EM axis uncertainty, for ET and CE ($\sigma_\epsilon=0^\circ$) and ET+CE ($\sigma_\epsilon=0^\circ$, $5^\circ$, $10^\circ$). The gray dashed curve shows the fraction with $\mathcal{S}_{\rm cm}\geq3$ (Eq.~\eqref{eq:scm}) for $\sigma_\epsilon=5^\circ$. The vertical dashed line marks the uncertainty in the jet position angle measured by VLBI for GW170817 at 40\,Mpc.}
\label{fig:results}
\end{figure*}

%% =====================================================================
\section{Results}
\label{sec:results}

\subsection{Image positions and arrival times (V1)}
\label{sec:v1}

Figs.~\ref{fig:results}(a) and (b) illustrate the prediction of later images for a lens with a fixed model, including a minor mirror solution~\cite{Hannuksela2020}. For the population, we include lens-model errors of 1\% on $\theta_E$, 0.05 on $\gamma'$, and 0.01 on each shear component, and a 1\% prior on $H_0$. We also include two selection terms: each posterior sample must produce the observed number of images, each above the detection threshold, and the magnification $|\mu_1|$ of the first GW image from its amplitude and tidal redshift enters with its Fisher uncertainty. Without both terms, the predictions favor low magnifications, and only 49\% and 86\% of the true second-image positions fall inside the nominal 68\% and 95\% regions, mostly due to the detection conditioning. With both terms, the coverage is calibrated (Fig.~\ref{fig:results}(c)): the 68\% and 95\% regions contain 69\% and 96\% of the true delays and 71\% and 98\% of the true positions after the first GW image (248 events), and 67\% and 98\% and 64\% and 92\% after the second (64 events), when the median delay uncertainty drops from 26\% to 8\%. A host prior with a 1.5 times too small effective radius lowers the position coverage to 54\% and 90\%, which V1 can diagnose.

\subsection{Absolute magnifications (V2)}
\label{sec:v2}

A mass-sheet transformation $\kappa\to\lambda\kappa+(1-\lambda)$, together with an unobservable rescaling $\bm\beta\to\lambda\bm\beta$, preserves image positions and flux ratios but maps $\mu_j\to\mu_j/\lambda^2$ and the inter-image delays $\Delta t_{jk}\to\lambda\Delta t_{jk}$~\cite{Falco1985}. Hence, the predicted $D_L^{{\rm obs},j}$ and $\Delta t_{jk}$ both scale as $\lambda/H_0$, and unlike standardizable sources~\cite{Oguri2003,Birrer2022}, an external $H_0$ or interference effects~\cite{Cremonese2021} are needed. Since $A_j\to\lambda A_j$, the offsets $\Delta\chi_j$ are invariant and cannot break this degeneracy. The GW-only $1\sigma$ uncertainty in $\ln|\mu_j|$ has a median of 0.68 (a factor of two) for ET+CE (Fig.~\ref{fig:results}(d)); with a 10\% lens-model floor and a 1\% $H_0$ prior, the per-event $\sigma_\lambda$ has a median of 0.31 and is below 0.1 for only 4\% of the events (Fig.~\ref{fig:results}(e)).

\subsection{Orientation offsets (V3)}
\label{sec:v3}

The median $|\Delta\chi_j|$ is $24^\circ$ ($15^\circ$ for minima and $45^\circ$ for saddles; Fig.~\ref{fig:results}(f)), but $\sigma_\psi$ is the limiting factor (Fig.~\ref{fig:gw}(c)), i.e., below a viewing angle $\min(\iota,180^\circ-\iota)\approx35^\circ$, $\sigma_\psi$ exceeds the uninformative value $90^\circ/\sqrt{12}\simeq26^\circ$ for $\psi$ defined modulo $90^\circ$ (Sec.~\ref{sec:asymmetry}). With ET+CE and $\sigma_\epsilon=0$, at least one image reaches $\mathcal{S}_j\geq3$ in 33\%, 25\%, and 10\% of the events for $\sigma_\chi=0^\circ$, $10^\circ$, and $20^\circ$ (Fig.~\ref{fig:results}(g)), all with viewing angles above $40^\circ$ and a median $\sigma_\psi$ of $7^\circ$, small enough to resolve the $90^\circ$ ambiguity. The common mode reaches $\mathcal{S}_{\rm cm}\geq3$ in 14\% of the events ($\sigma_\chi\to0$, $\sigma_\epsilon=5^\circ$).

%% =====================================================================
\section{Discussions}
\label{sec:discussion}

In strong lensing, the intrinsic orientation of a resolved structure is not directly observable and can be inferred only through a lens model, so EM observations alone directly measure only the relative orientations between images. In this work, we propose the GW polarization, which the lens leaves unchanged, as the missing orientation reference; its offset from the apparent axis of a resolved jet, such as that of GW170817~\cite{Mooley2018}, measures the lens-induced orientation offset of each image in absolute terms and, together with the image positions and Fermat-potential heights, completes the representation of lensed GWs on the EM image plane.

Our mock population supports each layer. The positions and arrival times of later images are predicted with calibrated uncertainties once the detection selection is included (V1), whereas GW-only absolute magnifications remain limited by the mass-sheet degeneracy (V2). The orientation offsets have a median of $24^\circ$, and at least one image reaches $\mathcal{S}_j\geq3$ in a third of the events for $\sigma_\chi\to0$ (V3). Since these offsets depend only on the reduced shear, they are invariant under the mass-sheet transformation and test the local lens structure at each image independently of the time delays; a mismatch would indicate lens-model errors, e.g., due to substructure, or non-parallel transport of the GW polarization.

These results are limited by the inputs rather than by the method. First, for quadrupole-dominated signals, $\psi$ is measured only at viewing angles above about $40^\circ$ (Fig.~\ref{fig:gw}(c)), whereas jets are most easily detected at smaller viewing angles. Second, EM inputs are rare, e.g., about $0.5\,{\rm yr}^{-1}$ lensed kilonovae for ET+CE with Roman-like follow-up, and host identification for 15\% (ET) and 30\% (CE) of events~\cite{ChenLuZeng2026}, and the GW170817 jet~\cite{Mooley2018} would appear 44 times smaller at $z_s=2$, three times smaller again in observer-frame time, and $\sim10^5$ times fainter after a $K$-correction for $F_\nu\propto\nu^{-0.6}$, so that V3 with BNS jets is realistic only for nearby, inclined systems. Third, GW170817 supports jet--orbit alignment only between its EM viewing angle and GW inclination~\cite{Finstad2018,Mooley2018}, and a scatter $\sigma_\epsilon=10^\circ$ reduces the V3 fractions by about a quarter (from 33\% to 26\% for $\sigma_\chi\to0$). Finally, we neglect non-Gaussian posteriors, angular structure of the lens, and microlensing by stars, which can distort highly magnified images~\cite{Mishra2021}.

These limitations call for follow-up work. Higher harmonics and precession break the distance--inclination degeneracy~\cite{Usman2019} and identify image types~\cite{Janquart2021}, and can thus measure $\psi$ at smaller viewing angles, where jets are brighter; exploiting them requires a full Bayesian analysis of the lensed images~\cite{Lo2023}. Lensed black hole binaries~\cite{Hannuksela2020,Yu2020,Chen2025}, with stronger higher harmonics, and subhalo-lensed LISA binaries~\cite{Caliskan2023}, with VLBI imaging of their hosts, would extend the framework beyond BNS mergers, which we will address in future work.

\section*{Data availability}
No data was used for the research described in the article

\section*{Declaration of competing interest}
The author declares that there are no known competing financial interests or personal relationships that could have appeared to influence the work reported in this paper.

\section*{Acknowledgments}
This work is supported by the Korea Astronomy and Space Science Institute under the R\&D program (Project No.~2026-1-810-00) supervised by the Ministry of Science and ICT of the Korea Government.

\bibliographystyle{elsarticle-num}
\bibliography{references}

@book{Schneider1992,
  author    = {Schneider, P. and Ehlers, J. and Falco, E. E.},
  title     = {Gravitational Lenses},
  publisher = {Springer-Verlag},
  address   = {Berlin},
  year      = {1992},
  doi = {10.1007/978-3-662-03758-4}
}

@article{Blandford1986,
  author  = {Blandford, R. and Narayan, R.},
  title   = {Fermat's principle, caustics, and the classification of gravitational lens images},
  journal = {Astrophys. J.},
  volume  = {310},
  pages   = {568},
  year    = {1986},
  doi     = {10.1086/164709}
}

@article{Treu2010,
  author  = {Treu, T.},
  title   = {Strong Lensing by Galaxies},
  journal = {Annu. Rev. Astron. Astrophys.},
  volume  = {48},
  pages   = {87},
  year    = {2010},
  doi     = {10.1146/annurev-astro-081309-130924}
}

@article{Falco1985,
  author  = {Falco, E. E. and Gorenstein, M. V. and Shapiro, I. I.},
  title   = {On model-dependent bounds on {$H_0$} from gravitational images: application to {Q0957+561A,B}},
  journal = {Astrophys. J. Lett.},
  volume  = {289},
  pages   = {L1},
  year    = {1985},
  doi     = {10.1086/184422}
}

@article{Oguri2003,
  author  = {Oguri, M. and Kawano, Y.},
  title   = {Gravitational lens time delays for distant supernovae: break the degeneracy between radial mass profiles and the {Hubble} constant},
  journal = {Mon. Not. R. Astron. Soc.},
  volume  = {338},
  pages   = {L25},
  year    = {2003},
  doi = {10.1046/j.1365-8711.2003.06290.x}
}

@article{Birrer2022,
  author  = {Birrer, S. and Dhawan, S. and Shajib, A. J.},
  title   = {The {Hubble} constant from strongly lensed supernovae with standardizable magnifications},
  journal = {Astrophys. J.},
  volume  = {924},
  pages   = {2},
  year    = {2022},
  doi = {10.3847/1538-4357/ac323a}
}

@article{Ohanian1974,
  author  = {Ohanian, H. C.},
  title   = {On the focusing of gravitational radiation},
  journal = {Int. J. Theor. Phys.},
  volume  = {9},
  pages   = {425},
  year    = {1974},
  doi     = {10.1007/BF01810927}
}

@article{Deguchi1986,
  author  = {Deguchi, S. and Watson, W. D.},
  title   = {Diffraction in gravitational lensing for compact objects of low mass},
  journal = {Astrophys. J.},
  volume  = {307},
  pages   = {30},
  year    = {1986},
  doi     = {10.1086/164389}
}

@article{Wang1996,
  author  = {Wang, Y. and Stebbins, A. and Turner, E. L.},
  title   = {Gravitational lensing of gravitational waves from merging neutron star binaries},
  journal = {Phys. Rev. Lett.},
  volume  = {77},
  pages   = {2875},
  year    = {1996},
  doi = {10.1103/PhysRevLett.77.2875}
}

@article{Nakamura1998,
  author  = {Nakamura, T. T.},
  title   = {Gravitational lensing of gravitational waves from inspiraling binaries by a point mass lens},
  journal = {Phys. Rev. Lett.},
  volume  = {80},
  pages   = {1138},
  year    = {1998},
  doi = {10.1103/PhysRevLett.80.1138}
}

@article{Takahashi2003,
  author  = {Takahashi, R. and Nakamura, T.},
  title   = {Wave effects in the gravitational lensing of gravitational waves from chirping binaries},
  journal = {Astrophys. J.},
  volume  = {595},
  pages   = {1039},
  year    = {2003},
  doi = {10.1086/377430}
}

@article{Dai2017,
  author  = {Dai, L. and Venumadhav, T.},
  title   = {On the waveforms of gravitationally lensed gravitational waves},
  journal = {arXiv e-prints},
  eprint  = {1702.04724},
  archivePrefix = {arXiv},
  year    = {2017},
  doi = {10.48550/arXiv.1702.04724},
  url = {https://arxiv.org/abs/1702.04724}
}

@article{Ezquiaga2021,
  author  = {Ezquiaga, J. M. and Holz, D. E. and Hu, W. and Lagos, M. and Wald, R. M.},
  title   = {Phase effects from strong gravitational lensing of gravitational waves},
  journal = {Phys. Rev. D},
  volume  = {103},
  pages   = {064047},
  year    = {2021},
  doi = {10.1103/PhysRevD.103.064047}
}

@article{Cremonese2021,
  author  = {Cremonese, P. and Ezquiaga, J. M. and Salzano, V.},
  title   = {Breaking the mass-sheet degeneracy with gravitational wave interference in lensed events},
  journal = {Phys. Rev. D},
  volume  = {104},
  pages   = {023503},
  year    = {2021},
  doi     = {10.1103/PhysRevD.104.023503}
}

@article{Ng2018,
  author  = {Ng, K. K. Y. and Wong, K. W. K. and Broadhurst, T. and Li, T. G. F.},
  title   = {Precise {LIGO} lensing rate predictions for binary black holes},
  journal = {Phys. Rev. D},
  volume  = {97},
  pages   = {023012},
  year    = {2018},
  doi     = {10.1103/PhysRevD.97.023012}
}

@article{Li2018,
  author  = {Li, S.-S. and Mao, S. and Zhao, Y. and Lu, Y.},
  title   = {Gravitational lensing of gravitational waves: a statistical perspective},
  journal = {Mon. Not. R. Astron. Soc.},
  volume  = {476},
  pages   = {2220},
  year    = {2018},
  doi     = {10.1093/mnras/sty411}
}

@article{Oguri2018,
  author  = {Oguri, M.},
  title   = {Effect of gravitational lensing on the distribution of gravitational waves from distant binary black hole mergers},
  journal = {Mon. Not. R. Astron. Soc.},
  volume  = {480},
  pages   = {3842},
  year    = {2018},
  doi     = {10.1093/mnras/sty2145}
}

@article{Piorkowska2013,
  author        = {Pi{\'o}rkowska, A. and Biesiada, M. and Zhu, Z.-H.},
  title         = {Strong gravitational lensing of gravitational waves in {Einstein Telescope}},
  journal       = {J. Cosmol. Astropart. Phys.},
  volume        = {2013},
  number        = {10},
  pages         = {022},
  year          = {2013},
  doi           = {10.1088/1475-7516/2013/10/022},
  eprint        = {1309.5731},
  archivePrefix = {arXiv}
}

@article{Biesiada2014,
  author        = {Biesiada, M. and Ding, X. and Pi{\'o}rkowska, A. and Zhu, Z.-H.},
  title         = {Strong gravitational lensing of gravitational waves from double compact binaries---perspectives for the {Einstein Telescope}},
  journal       = {J. Cosmol. Astropart. Phys.},
  volume        = {2014},
  number        = {10},
  pages         = {080},
  year          = {2014},
  doi           = {10.1088/1475-7516/2014/10/080},
  eprint        = {1409.8360},
  archivePrefix = {arXiv}
}

@article{Ding2015,
  author        = {Ding, X. and Biesiada, M. and Zhu, Z.-H.},
  title         = {Strongly lensed gravitational waves from intrinsically faint double compact binaries---prediction for the {Einstein Telescope}},
  journal       = {J. Cosmol. Astropart. Phys.},
  volume        = {2015},
  number        = {12},
  pages         = {006},
  year          = {2015},
  doi           = {10.1088/1475-7516/2015/12/006},
  eprint        = {1508.05000},
  archivePrefix = {arXiv}
}

@article{Wierda2021,
  author        = {Wierda, A. R. A. C. and Wempe, E. and Hannuksela, O. A. and Koopmans, L. V. E. and Van Den Broeck, C.},
  title         = {Beyond the detector horizon: forecasting gravitational-wave strong lensing},
  journal       = {Astrophys. J.},
  volume        = {921},
  number        = {2},
  pages         = {154},
  year          = {2021},
  doi           = {10.3847/1538-4357/ac1bb4},
  eprint        = {2106.06303},
  archivePrefix = {arXiv}
}

@article{Xu2022,
  author        = {Xu, F. and Ezquiaga, J. M. and Holz, D. E.},
  title         = {Please repeat: strong lensing of gravitational waves as a probe of compact binary and galaxy populations},
  journal       = {Astrophys. J.},
  volume        = {929},
  number        = {1},
  pages         = {9},
  year          = {2022},
  doi           = {10.3847/1538-4357/ac58f8},
  eprint        = {2105.14390},
  archivePrefix = {arXiv}
}

@article{Hannuksela2019,
  author        = {Hannuksela, O. A. and Haris, K. and Ng, K. K. Y. and Kumar, S. and Mehta, A. K. and Keitel, D. and Li, T. G. F. and Ajith, P.},
  title         = {Search for gravitational lensing signatures in {LIGO}--{Virgo} binary black hole events},
  journal       = {Astrophys. J. Lett.},
  volume        = {874},
  number        = {1},
  pages         = {L2},
  year          = {2019},
  doi           = {10.3847/2041-8213/ab0c0f},
  eprint        = {1901.02674},
  archivePrefix = {arXiv}
}

@article{LVK2025lensing,
  author        = {Abac, A. G. and others},
  collaboration = {LIGO Scientific Collaboration and Virgo Collaboration and KAGRA Collaboration},
  title         = {{GWTC-4.0}: Searches for gravitational-wave lensing signatures},
  journal       = {Astrophys. J.},
  volume        = {1011},
  number        = {2},
  pages         = {240},
  year          = {2026},
  doi           = {10.3847/1538-4357/ae9744},
  eprint        = {2512.16347},
  archivePrefix = {arXiv}
}

@article{LVK2021lensing,
  author        = {Abbott, R. and others},
  collaboration = {LIGO Scientific Collaboration and Virgo Collaboration},
  title         = {Search for lensing signatures in the gravitational-wave observations from the first half of {LIGO}--{Virgo}'s third observing run},
  journal       = {Astrophys. J.},
  volume        = {923},
  number        = {1},
  pages         = {14},
  year          = {2021},
  doi           = {10.3847/1538-4357/ac23db},
  eprint        = {2105.06384},
  archivePrefix = {arXiv}
}

@article{LVK2023lensing,
  author        = {Abbott, R. and others},
  collaboration = {LIGO Scientific Collaboration and Virgo Collaboration and KAGRA Collaboration},
  title         = {Search for gravitational-lensing signatures in the full third observing run of the {LIGO}--{Virgo} network},
  journal       = {Astrophys. J.},
  volume        = {970},
  number        = {2},
  pages         = {191},
  year          = {2024},
  doi           = {10.3847/1538-4357/ad3e83},
  eprint        = {2304.08393},
  archivePrefix = {arXiv}
}

@article{Bartelmann2001,
  author        = {Bartelmann, M. and Schneider, P.},
  title         = {Weak gravitational lensing},
  journal       = {Phys. Rep.},
  volume        = {340},
  number        = {4--5},
  pages         = {291--472},
  year          = {2001},
  doi           = {10.1016/S0370-1573(00)00082-X},
  eprint        = {astro-ph/9912508},
  archivePrefix = {arXiv}
}

@article{Ronchini2022,
  author        = {Ronchini, S. and Branchesi, M. and Oganesyan, G. and Banerjee, B. and Dupletsa, U. and Ghirlanda, G. and Harms, J. and Mapelli, M. and Santoliquido, F.},
  title         = {Perspectives for multimessenger astronomy with the next generation of gravitational-wave detectors and high-energy satellites},
  journal       = {Astron. Astrophys.},
  volume        = {665},
  pages         = {A97},
  year          = {2022},
  doi           = {10.1051/0004-6361/202243705},
  eprint        = {2204.01746},
  archivePrefix = {arXiv}
}

@article{Bayer2020,
  author        = {Bayer, A. E. and Seljak, U.},
  title         = {The look-elsewhere effect from a unified {Bayesian} and frequentist perspective},
  journal       = {J. Cosmol. Astropart. Phys.},
  volume        = {2020},
  number        = {10},
  pages         = {009},
  year          = {2020},
  doi           = {10.1088/1475-7516/2020/10/009},
  eprint        = {2007.13821},
  archivePrefix = {arXiv}
}

@article{Caliskan2023luck,
  author        = {{\c{C}}al{\i}{\c{s}}kan, M. and Ezquiaga, J. M. and Hannuksela, O. A. and Holz, D. E.},
  title         = {Lensing or luck? {False} alarm probabilities for gravitational lensing of gravitational waves},
  journal       = {Phys. Rev. D},
  volume        = {107},
  pages         = {063023},
  year          = {2023},
  doi           = {10.1103/PhysRevD.107.063023},
  eprint        = {2201.04619},
  archivePrefix = {arXiv}
}

@article{Planck2018,
  author        = {Aghanim, N. and others},
  collaboration = {Planck Collaboration},
  title         = {{Planck} 2018 results. {VI}. {Cosmological} parameters},
  journal       = {Astron. Astrophys.},
  volume        = {641},
  pages         = {A6},
  year          = {2020},
  doi           = {10.1051/0004-6361/201833910},
  eprint        = {1807.06209},
  archivePrefix = {arXiv}
}

@article{LVC2018EOS,
  author        = {Abbott, B. P. and others},
  collaboration = {LIGO Scientific Collaboration and Virgo Collaboration},
  title         = {{GW170817}: Measurements of neutron star radii and equation of state},
  journal       = {Phys. Rev. Lett.},
  volume        = {121},
  number        = {16},
  pages         = {161101},
  year          = {2018},
  doi           = {10.1103/PhysRevLett.121.161101},
  eprint        = {1805.11581},
  archivePrefix = {arXiv}
}

@article{De2018,
  author        = {De, S. and Finstad, D. and Lattimer, J. M. and Brown, D. A. and Berger, E. and Biwer, C. M.},
  title         = {Tidal deformabilities and radii of neutron stars from the observation of {GW170817}},
  journal       = {Phys. Rev. Lett.},
  volume        = {121},
  number        = {9},
  pages         = {091102},
  year          = {2018},
  doi           = {10.1103/PhysRevLett.121.091102},
  eprint        = {1804.08583},
  archivePrefix = {arXiv}
}

@article{Punturo2010,
  author        = {Punturo, M. and others},
  title         = {The {Einstein Telescope}: a third-generation gravitational wave observatory},
  journal       = {Class. Quantum Grav.},
  volume        = {27},
  number        = {19},
  pages         = {194002},
  year          = {2010},
  doi           = {10.1088/0264-9381/27/19/194002}
}

@article{Reitze2019,
  author        = {Reitze, D. and Adhikari, R. X. and Ballmer, S. and Barish, B. and others},
  title         = {{Cosmic Explorer}: the {U.S.} contribution to gravitational-wave astronomy beyond {LIGO}},
  journal       = {Bull. Am. Astron. Soc.},
  volume        = {51},
  number        = {7},
  pages         = {035},
  year          = {2019},
  doi           = {10.48550/arXiv.1907.04833},
  eprint        = {1907.04833},
  archivePrefix = {arXiv}
}

@article{ETScience2026,
  author        = {Abac, A. and Abramo, R. and Albanesi, S. and others},
  title         = {The science of the {Einstein Telescope}},
  journal       = {J. Cosmol. Astropart. Phys.},
  volume        = {2026},
  number        = {03},
  pages         = {081},
  year          = {2026},
  doi           = {10.1088/1475-7516/2026/03/081},
  eprint        = {2503.12263},
  archivePrefix = {arXiv}
}

@article{Maggiore2020,
  author  = {Maggiore, M. and others},
  title   = {Science case for the {Einstein} telescope},
  journal = {J. Cosmol. Astropart. Phys.},
  volume  = {2020},
  number  = {03},
  pages   = {050},
  year    = {2020},
  doi = {10.1088/1475-7516/2020/03/050}
}

@article{Branchesi2023,
  author  = {Branchesi, M. and others},
  title   = {Science with the {Einstein} Telescope: a comparison of different designs},
  journal = {J. Cosmol. Astropart. Phys.},
  volume  = {2023},
  number  = {07},
  pages   = {068},
  year    = {2023},
  doi = {10.1088/1475-7516/2023/07/068}
}

@article{Evans2021,
  author  = {Evans, M. and others},
  title   = {A horizon study for {Cosmic Explorer}: science, observatories, and community},
  journal = {arXiv e-prints},
  eprint  = {2109.09882},
  archivePrefix = {arXiv},
  year    = {2021},
  doi = {10.48550/arXiv.2109.09882},
  url = {https://arxiv.org/abs/2109.09882}
}

@article{Seto2004,
  author  = {Seto, N.},
  title   = {Strong gravitational lensing and localization of merging massive black hole binaries with {LISA}},
  journal = {Phys. Rev. D},
  volume  = {69},
  pages   = {022002},
  year    = {2004},
  eprint  = {astro-ph/0305605},
  archivePrefix = {arXiv},
  doi = {10.1103/PhysRevD.69.022002}
}

@article{Hannuksela2020,
  author  = {Hannuksela, O. A. and Collett, T. E. and {\c{C}}al{\i}{\c{s}}kan, M. and Li, T. G. F.},
  title   = {Localizing merging black holes with sub-arcsecond precision using gravitational-wave lensing},
  journal = {Mon. Not. R. Astron. Soc.},
  volume  = {498},
  pages   = {3395},
  year    = {2020},
  eprint  = {2004.13811},
  archivePrefix = {arXiv},
  doi = {10.1093/mnras/staa2577}
}

@article{Yu2020,
  author  = {Yu, H. and Zhang, P. and Wang, F.-Y.},
  title   = {Strong lensing as a giant telescope to localize the host galaxy of gravitational wave event},
  journal = {Mon. Not. R. Astron. Soc.},
  volume  = {497},
  pages   = {204},
  year    = {2020},
  eprint  = {2007.00828},
  archivePrefix = {arXiv},
  doi = {10.1093/mnras/staa1952}
}

@article{Chen2025,
  author  = {Chen, Z. and Yu, Q. and Lu, Y. and Guo, X.},
  title   = {Enhanced localization of dark lensed gravitational wave events enables host galaxy identification and precise cosmological inference},
  journal = {arXiv e-prints},
  eprint  = {2510.12470},
  archivePrefix = {arXiv},
  year    = {2025},
  doi = {10.48550/arXiv.2510.12470},
  url = {https://arxiv.org/abs/2510.12470}
}

@article{Liao2017,
  author  = {Liao, K. and Fan, X.-L. and Ding, X. and Biesiada, M. and Zhu, Z.-H.},
  title   = {Precision cosmology from future lensed gravitational wave and electromagnetic signals},
  journal = {Nat. Commun.},
  volume  = {8},
  pages   = {1148},
  year    = {2017},
  doi = {10.1038/s41467-017-01152-9}
}

@article{Collett2017,
  author  = {Collett, T. E. and Bacon, D.},
  title   = {Testing the speed of gravitational waves over cosmological distances with strong gravitational lensing},
  journal = {Phys. Rev. Lett.},
  volume  = {118},
  pages   = {091101},
  year    = {2017},
  doi = {10.1103/PhysRevLett.118.091101}
}

@article{Fan2017,
  author  = {Fan, X.-L. and Liao, K. and Biesiada, M. and Pi{\'o}rkowska-Kurpas, A. and Zhu, Z.-H.},
  title   = {Speed of gravitational waves from strongly lensed gravitational waves and electromagnetic signals},
  journal = {Phys. Rev. Lett.},
  volume  = {118},
  pages   = {091102},
  year    = {2017},
  doi = {10.1103/PhysRevLett.118.091102}
}

@article{Yang2019,
  author  = {Yang, T. and Hu, B. and Cai, R.-G. and Wang, B.},
  title   = {New probe of gravity: strongly lensed gravitational wave multimessenger approach},
  journal = {Astrophys. J.},
  volume  = {880},
  pages   = {50},
  year    = {2019},
  eprint  = {1810.00164},
  archivePrefix = {arXiv},
  doi = {10.3847/1538-4357/ab271e}
}

@article{ChenLuZeng2026,
  author  = {Chen, Z. and Lu, Y. and Zeng, C.},
  title   = {Prospect for detection of strongly lensed multi-messenger signals of binary neutron star mergers},
  journal = {arXiv e-prints},
  eprint  = {2607.07120},
  archivePrefix = {arXiv},
  year    = {2026},
  doi = {10.48550/arXiv.2607.07120},
  url = {https://arxiv.org/abs/2607.07120}
}

@article{Smith2025,
  author        = {Smith, G. P. and Baker, T. and Birrer, S. and Collins, C. E. and Ezquiaga, J. M. and Goyal, S. and Hannuksela, O. A. and Hemanta, P. and Hendry, M. A. and Janquart, J. and others},
  title         = {Multi-messenger gravitational lensing},
  journal       = {Phil. Trans. R. Soc. A},
  volume        = {383},
  number        = {2295},
  pages         = {20240134},
  year          = {2025},
  doi           = {10.1098/rsta.2024.0134},
  eprint        = {2503.19973},
  archivePrefix = {arXiv}
}

@article{Ryczanowski2025,
  author  = {Ryczanowski, D. and Cooke, J. and Freeburn, J. and Gompertz, B. and Haines, C. P. and Nicholl, M. and Smith, G. P. and {Van Bemmel}, N. and Zhang, J.},
  title   = {A follow-up strategy enabling discovery of electromagnetic counterparts to highly magnified gravitationally lensed gravitational waves},
  journal = {Phil. Trans. R. Soc. A},
  eprint  = {2503.21811},
  archivePrefix = {arXiv},
  year    = {2025},
  volume = {383},
  number = {2295},
  pages = {20240118},
  doi = {10.1098/rsta.2024.0118}
}

@article{Magare2023,
  author  = {Magare, S. and Kapadia, S. J. and More, A. and Singh, M. K. and Ajith, P. and Ramprakash, A. N.},
  title   = {Gear up for the action replay: leveraging lensing for enhanced gravitational-wave early warning},
  journal = {Astrophys. J. Lett.},
  eprint  = {2302.02916},
  archivePrefix = {arXiv},
  year    = {2023},
  note    = {Article number: CHECK},
  volume = {955},
  number = {2},
  doi = {10.3847/2041-8213/acf668}
}

@article{Hou2019,
  author  = {Hou, S. and Fan, X.-L. and Zhu, Z.-H.},
  title   = {Gravitational lensing of gravitational waves: rotation of polarization plane},
  journal = {Phys. Rev. D},
  volume  = {100},
  pages   = {064028},
  year    = {2019},
  eprint  = {1907.07486},
  archivePrefix = {arXiv},
  note    = {Journal reference: CHECK},
  doi = {10.1103/PhysRevD.100.064028}
}

@article{Goyal2021,
  author  = {Goyal, S. and Haris, K. and Mehta, A. K. and Ajith, P.},
  title   = {Testing the nature of gravitational-wave polarizations using strongly lensed signals},
  journal = {Phys. Rev. D},
  volume  = {103},
  pages   = {024038},
  year    = {2021},
  eprint  = {2008.07060},
  archivePrefix = {arXiv},
  note    = {Author list: CHECK},
  doi = {10.1103/PhysRevD.103.024038}
}

@article{MaganaHernandez2022,
  author  = {{Maga{\~n}a Hernandez}, I.},
  title   = {Measuring the polarization content of gravitational waves with strongly lensed binary black hole mergers},
  journal = {arXiv e-prints},
  eprint  = {2211.01272},
  archivePrefix = {arXiv},
  year    = {2022},
  doi = {10.48550/arXiv.2211.01272},
  url = {https://arxiv.org/abs/2211.01272}
}

@article{Andersson2023,
  author  = {Andersson, L. and Oancea, M. A.},
  title   = {Spin {Hall} effects in the sky},
  journal = {Class. Quantum Grav.},
  eprint  = {2302.13634},
  archivePrefix = {arXiv},
  year    = {2023},
  volume = {40},
  pages = {154002},
  doi = {10.1088/1361-6382/ace021}
}

@article{Imam2026,
  author  = {Imam, S. M. A. and Lagos, M.},
  title   = {Tests of scalar polarizations with multi-messenger events},
  journal = {Phys. Rev. D},
  eprint  = {2604.24526},
  archivePrefix = {arXiv},
  year    = {2026},
  doi = {10.1103/f9ll-y1rn},
  note = {Volume and article number: CHECK}
}

@article{Surpi1999,
  author  = {Surpi, G. C. and Harari, D. D.},
  title   = {Weak lensing by large-scale structure and the polarization properties of distant radio sources},
  journal = {Astrophys. J.},
  volume  = {515},
  pages   = {455},
  year    = {1999},
  eprint  = {astro-ph/9709087},
  archivePrefix = {arXiv},
  doi = {10.1086/307031}
}

@article{Brown2011,
  author  = {Brown, M. L. and Battye, R. A.},
  title   = {Polarization as an indicator of intrinsic alignment in radio weak lensing},
  journal = {Mon. Not. R. Astron. Soc.},
  volume  = {410},
  pages   = {2057},
  year    = {2011},
  eprint  = {1005.1926},
  archivePrefix = {arXiv},
  doi = {10.1111/j.1365-2966.2010.17583.x}
}

@article{Francfort2021,
  author  = {Francfort, J. and Cusin, G. and Durrer, R.},
  title   = {Image rotation from weak lensing},
  journal = {Class. Quantum Grav.},
  eprint  = {2106.08631},
  archivePrefix = {arXiv},
  year    = {2021},
  doi = {10.1088/1361-6382/ac33ba},
  note = {Volume and article number: CHECK}
}

@article{GW170817,
  author        = {Abbott, B. P. and others},
  collaboration = {LIGO Scientific Collaboration and Virgo Collaboration},
  title         = {{GW170817}: observation of gravitational waves from a binary neutron star inspiral},
  journal       = {Phys. Rev. Lett.},
  volume        = {119},
  pages         = {161101},
  year          = {2017},
  doi           = {10.1103/PhysRevLett.119.161101},
  eprint        = {1710.05832},
  archivePrefix = {arXiv}
}

@article{GW170817MMA,
  author        = {Abbott, B. P. and others},
  collaboration = {LIGO Scientific Collaboration and Virgo Collaboration and others},
  title         = {Multi-messenger observations of a binary neutron star merger},
  journal       = {Astrophys. J. Lett.},
  volume        = {848},
  number        = {2},
  pages         = {L12},
  year          = {2017},
  doi           = {10.3847/2041-8213/aa91c9},
  eprint        = {1710.05833},
  archivePrefix = {arXiv}
}

@article{Mooley2018,
  author  = {Mooley, K. P. and others},
  title   = {Superluminal motion of a relativistic jet in the neutron-star merger {GW170817}},
  journal = {Nature},
  volume  = {561},
  pages   = {355},
  year    = {2018},
  doi = {10.1038/s41586-018-0486-3}
}

@article{Ghirlanda2019,
  author  = {Ghirlanda, G. and others},
  title   = {Compact radio emission indicates a structured jet was produced by a binary neutron star merger},
  journal = {Science},
  volume  = {363},
  pages   = {968},
  year    = {2019},
  doi = {10.1126/science.aau8815}
}

@article{Hotokezaka2019,
  author  = {Hotokezaka, K. and Nakar, E. and Gottlieb, O. and Nissanke, S. and Masuda, K. and Hallinan, G. and Mooley, K. P. and Deller, A. T.},
  title   = {A {Hubble} constant measurement from superluminal motion of the jet in {GW170817}},
  journal = {Nat. Astron.},
  volume  = {3},
  pages   = {940},
  year    = {2019},
  doi = {10.1038/s41550-019-0820-1}
}

@article{Messenger2012,
  author  = {Messenger, C. and Read, J.},
  title   = {Measuring a cosmological distance-redshift relationship using only gravitational wave observations of binary neutron star coalescences},
  journal = {Phys. Rev. Lett.},
  volume  = {108},
  pages   = {091101},
  year    = {2012},
  doi = {10.1103/PhysRevLett.108.091101}
}

@article{DelPozzo2017,
  author  = {{Del Pozzo}, W. and Li, T. G. F. and Messenger, C.},
  title   = {Cosmological inference using only gravitational wave observations of binary neutron stars},
  journal = {Phys. Rev. D},
  volume  = {95},
  pages   = {043502},
  year    = {2017},
  doi = {10.1103/PhysRevD.95.043502}
}

@article{Caliskan2023,
  author  = {{\c{C}}al{\i}{\c{s}}kan, M. and Kumar, N. A. and Ji, L. and Ezquiaga, J. M. and Cotesta, R. and Berti, E. and Kamionkowski, M.},
  title   = {Probing wave-optics effects and low-mass dark matter halos with lensing of gravitational waves from massive black holes},
  journal = {Phys. Rev. D},
  volume  = {108},
  pages   = {123543},
  year    = {2023},
  eprint  = {2307.06990},
  archivePrefix = {arXiv},
  doi = {10.1103/PhysRevD.108.123543}
}

@article{Arun2005,
  author  = {Arun, K. G. and Iyer, B. R. and Sathyaprakash, B. S. and Sundararajan, P. A.},
  title   = {Parameter estimation of inspiralling compact binaries using 3.5 post-{Newtonian} gravitational wave phasing: the nonspinning case},
  journal = {Phys. Rev. D},
  volume  = {71},
  pages   = {084008},
  year    = {2005},
  doi = {10.1103/PhysRevD.71.084008}
}

@article{Flanagan2008,
  author  = {Flanagan, {\'E}. {\'E}. and Hinderer, T.},
  title   = {Constraining neutron-star tidal {Love} numbers with gravitational-wave detectors},
  journal = {Phys. Rev. D},
  volume  = {77},
  pages   = {021502},
  year    = {2008},
  doi = {10.1103/PhysRevD.77.021502}
}

@article{Wade2014,
  author  = {Wade, L. and Creighton, J. D. E. and Ochsner, E. and Lackey, B. D. and Farr, B. F. and Littenberg, T. B. and Raymond, V.},
  title   = {Systematic and statistical errors in a {Bayesian} approach to the estimation of the neutron-star equation of state using advanced gravitational wave detectors},
  journal = {Phys. Rev. D},
  volume  = {89},
  pages   = {103012},
  year    = {2014},
  doi = {10.1103/PhysRevD.89.103012}
}

@article{Hild2011,
  author  = {Hild, S. and others},
  title   = {Sensitivity studies for third-generation gravitational wave observatories},
  journal = {Class. Quantum Grav.},
  volume  = {28},
  pages   = {094013},
  year    = {2011},
  doi = {10.1088/0264-9381/28/9/094013},
  number = {9}
}

@article{Abbott2017CE,
  author        = {Abbott, B. P. and others},
  collaboration = {LIGO Scientific Collaboration},
  title         = {Exploring the sensitivity of next generation gravitational wave detectors},
  journal       = {Class. Quantum Grav.},
  volume        = {34},
  number        = {4},
  pages         = {044001},
  year          = {2017},
  doi           = {10.1088/1361-6382/aa51f4},
  eprint        = {1607.08697},
  archivePrefix = {arXiv}
}

@article{Ashton2019,
  author  = {Ashton, G. and others},
  title   = {{BILBY}: a user-friendly {Bayesian} inference library for gravitational-wave astronomy},
  journal = {Astrophys. J. Suppl.},
  volume  = {241},
  pages   = {27},
  year    = {2019},
  doi = {10.3847/1538-4365/ab06fc}
}

@article{Choi2007,
  author  = {Choi, Y.-Y. and Park, C. and Vogeley, M. S.},
  title   = {Internal and collective properties of galaxies in the {Sloan Digital Sky Survey}},
  journal = {Astrophys. J.},
  volume  = {658},
  pages   = {884},
  year    = {2007},
  doi = {10.1086/511060}
}

@article{Madau2014,
  author  = {Madau, P. and Dickinson, M.},
  title   = {Cosmic star-formation history},
  journal = {Annu. Rev. Astron. Astrophys.},
  volume  = {52},
  pages   = {415},
  year    = {2014},
  doi = {10.1146/annurev-astro-081811-125615}
}

@article{Turner1984,
  author  = {Turner, E. L. and Ostriker, J. P. and Gott, J. R.},
  title   = {The statistics of gravitational lenses: the distributions of image angular separations and lens redshifts},
  journal = {Astrophys. J.},
  volume  = {284},
  pages   = {1},
  year    = {1984},
  doi     = {10.1086/162379}
}

@article{Ozel2012,
  author  = {{\"O}zel, F. and Psaltis, D. and Narayan, R. and Santos Villarreal, A.},
  title   = {On the mass distribution and birth masses of neutron stars},
  journal = {Astrophys. J.},
  volume  = {757},
  pages   = {55},
  year    = {2012},
  doi = {10.1088/0004-637X/757/1/55},
  number = {1}
}

@article{Finstad2018,
  author  = {Finstad, D. and De, S. and Brown, D. A. and Berger, E. and Biwer, C. M.},
  title   = {Measuring the viewing angle of {GW170817} with electromagnetic and gravitational waves},
  journal = {Astrophys. J. Lett.},
  volume  = {860},
  pages   = {L2},
  year    = {2018},
  eprint  = {1804.04179},
  archivePrefix = {arXiv},
  note    = {Volume and page: CHECK},
  doi = {10.3847/2041-8213/aac6c1}
}

@article{Pagano2020,
  author        = {Pagano, G. and Hannuksela, O. A. and Li, T. G. F.},
  title         = {{lensingGW}: a {Python} package for lensing of gravitational waves},
  journal       = {Astron. Astrophys.},
  volume        = {643},
  pages         = {A167},
  year          = {2020},
  doi           = {10.1051/0004-6361/202038730},
  eprint        = {2006.12879},
  archivePrefix = {arXiv},
  note          = {Volume: CHECK}
}

@article{Wempe2024,
  author        = {Wempe, E. and Koopmans, L. V. E. and Wierda, A. R. A. C. and Hannuksela, O. A. and {Van Den Broeck}, C.},
  title         = {On the detection and precise localization of merging black holes events through strong gravitational lensing},
  journal       = {Mon. Not. R. Astron. Soc.},
  volume        = {530},
  number        = {3},
  pages         = {3368--3390},
  year          = {2024},
  doi           = {10.1093/mnras/stae1023},
  eprint        = {2204.08732},
  archivePrefix = {arXiv}
}

@article{Usman2019,
  author  = {Usman, S. A. and Mills, J. C. and Fairhurst, S.},
  title   = {Constraining the inclinations of binary mergers from gravitational-wave observations},
  journal = {Astrophys. J.},
  volume  = {877},
  pages   = {82},
  year    = {2019},
  doi     = {10.3847/1538-4357/ab0b3e}
}

@article{Janquart2021,
  author  = {Janquart, J. and Seo, E. and Hannuksela, O. A. and Li, T. G. F. and Van Den Broeck, C.},
  title   = {On the identification of individual gravitational-wave image types of a lensed system using higher-order modes},
  journal = {Astrophys. J. Lett.},
  volume  = {923},
  pages   = {L1},
  year    = {2021},
  doi     = {10.3847/2041-8213/ac3bcf}
}

@article{Mishra2021,
  author  = {Mishra, A. and Meena, A. K. and More, A. and Bose, S. and Bagla, J. S.},
  title   = {Gravitational lensing of gravitational waves: effect of microlens population in lensing galaxies},
  journal = {Mon. Not. R. Astron. Soc.},
  volume  = {508},
  pages   = {4869},
  year    = {2021},
  doi     = {10.1093/mnras/stab2875}
}

\end{document}